\documentclass[12pt]{article}

\ifx\pdfoutput\undefined
\usepackage[dvips,bookmarks]{hyperref}
\else
\usepackage{hyperref}
\fi
\hypersetup{colorlinks=false,bookmarksopen,bookmarksnumbered,citecolor=blue,
   pdfstartview=FitH}

\usepackage{latexsym}
\usepackage{amssymb,amsfonts,amsmath}
\usepackage{graphicx} 
\usepackage{indentfirst}
\usepackage{bbm}
\usepackage{amssymb}
\usepackage{verbatim}
\usepackage{amsmath, amsthm,amssymb}
\usepackage{mathrsfs}
\usepackage{hyperref}
\usepackage{amsfonts}
\usepackage{dsfont}
\usepackage{slashed, tensor}
\usepackage{booktabs}
\usepackage{graphicx}
\usepackage{mathrsfs}

\parskip=\medskipamount

\usepackage[mathscr]{euscript} 

\newcommand {\cC}{{\cal C}}
\newcommand {\cD}{{\cal D}}
\newcommand {\cE}{{\cal E}}

\newcommand {\cG}{{\cal G}}

\newcommand {\cK}{{\cal K}}

\newcommand {\cM}{{\cal M}}
\newcommand {\cN}{{\cal N}}

\newcommand {\cS}{{\cal S}}

\def\a{\alpha}
\def\b{\beta}

\def\d{\delta}

\def\f{\phi}
\def\g{\gamma}

\def\k{\kappa}

\def\m{\mu}

\def\q{\theta}

\def\s{\sigma}
\def\t{\tau}

\def\x{\xi}

\def\F{\Phi}

\def\O{\Omega}

\def\ri{{\rm i}}
\def\re{{\rm e}}

\newcommand{\gd}{{\dot\g}}

\newcommand{\ad}{{\dot{\alpha}}}
\newcommand{\bd}{{\dot{\beta}}}
\newcommand{\dalpha}{{\dot{\alpha}}}

\newcommand{\sSU}{\mathsf{SU}}
\newcommand{\sSL}{\mathsf{SL}}

\newcommand{\sU}{\mathsf{U}}

\newcommand{\ve}{\varepsilon}
\newcommand{\cDB}{{\bar\cD}}

\newcommand{\pa}{\partial}
\newcommand{\hf}{\frac12}

\newcommand{\be}{\begin{equation}}
\newcommand{\ee}{\end{equation}}
\newcommand{\bea}{\begin{eqnarray}}
\newcommand{\eea}{\end{eqnarray}}
\newcommand{\non}{\nonumber}
\newcommand{\ba}{\begin{array}}
\newcommand{\ea}{\end{array}}

\newcommand{\bm}[1]{\mbox{\boldmath$#1$}}

\def\double #1{#1{\hbox{\kern-2pt $#1$}}}

\newcommand{\CD}{{\nabla}}

\newcommand{\bsubeq}{\begin{subequations}}
\newcommand{\esubeq}{\end{subequations}}

\newcommand{\rd}{\mathrm d}
\numberwithin{equation}{section}  

\renewcommand{\CD}{{\nabla}}

\begin{document}

\begin{titlepage}
\begin{flushright}
June, 2016 \\
\end{flushright}
\vspace{5mm}

\begin{center}
{\Large \bf 
Maximally supersymmetric solutions  \\
of \mbox{$\bm{R^2}$}  supergravity}
\\ 
\end{center}

\begin{center}

{\bf
Sergei M. Kuzenko
} \\
\vspace{5mm}

\footnotesize{{\it School of Physics M013, The University of Western Australia\\
35 Stirling Highway, Crawley W.A. 6009, Australia}}  
\vspace{2mm}
~\\
\texttt{sergei.kuzenko@uwa.edu.au}\\
\vspace{2mm}

\end{center}

\begin{abstract}
\baselineskip=14pt
There are five  maximally supersymmetric backgrounds 
in four-dimensional off-shell $\cN=1$  supergravity, two of which are well known: 
Minkowski superspace ${\mathbb M}^{4|4}$ and 
anti-de Sitter superspace ${\rm AdS}^{4|4}$. 
The three remaining supermanifolds support spacetimes of different topology, 
which are: ${\mathbb R} \times S^3$, ${\rm AdS}_3 \times {\mathbb R}$, 
and a supersymmetric plane wave isometric to the Nappi-Witten group.  
As is well known, the Minkowski  and 
anti-de Sitter superspaces are solutions of the Poincar\'e  and anti-de Sitter 
supergravity theories, respectively. Here we demonstrate that the other three
superspaces are solutions of no-scale $R^2$ supergravity. 
We also present a new (probably the simplest) derivation of the maximally supersymmetric backgrounds of off-shell $\cN=1$  supergravity. 
\end{abstract}

\vfill

\vfill
\end{titlepage}

\newpage
\renewcommand{\thefootnote}{\arabic{footnote}}
\setcounter{footnote}{0}




\allowdisplaybreaks

\section{Introduction}

There exist only  five maximally supersymmetric backgrounds 
in off-shell $\cN=1$  supergravity in four dimensions. As purely bosonic backgrounds, 
the complete list was given by Festuccia and Seiberg \cite{FS}. 
Their results were re-derived in \cite{K13} using the superspace formalism
developed in the mid-1990s \cite{BK} (see \cite{K15Corfu} for a review).
As curved $\cN=1$ superspaces, all these backgrounds were described
in \cite{KT-M}. The algebraic aspects of these backgrounds have recently been studied in \cite{deMF-OS}. 

We now list all maximally supersymmetric backgrounds 
of $\cN=1$ supergravity following \cite{KT-M}.\footnote{In all cases, the superspace
covariant derivatives $\cD_A = (\cD_a, \cD_\a, \bar \cD^\ad)$ 
have the form
$\cD_{A}=(\cD_a,\cD_\a,{\bar \cD}^\ad)  
= E_A{}^M\pa_M+\hf \O_A{}^{bc}M_{bc}$,
where $M_{bc}$ is the Lorentz generator. In the case of Minkowski superspace, 
one can choose 
the Lorentz connection $\O_A{}^{bc}$ to vanish, 
and the inverse vielbein $E_A{}^M$ to have the Akulov-Volkov form \cite{AV}.} 
The simplest and most well-known is 
Minkowski superspace 
${\mathbb M}^{4|4}$  \cite{AV,SS}. 
It is characterised by the algebra of covariant derivatives
\begin{subequations} \label{Minkowski} 
\bea
& \{\cD_\a,\cDB_\bd\}=-2\ri\cD_{\a\bd}
~,\\ 
& 
\{\cD_\a,\cD_\b\}=0~,\quad
\{\cDB_\ad,\cDB_\bd\}=0~, \\
& {[}\cD_a,\cD_B{]}= 0~.
\eea
\end{subequations}
The second oldest background is 
anti-de Sitter (AdS) superspace ${\rm AdS}^{4|4}$ \cite{Keck,Zumino77,IS}.
It is characterised by the algebra of covariant derivatives
\begin{subequations}\label{AdS}
\bea
& \{\cD_\a,\cDB_\bd\}=-2\ri\cD_{\a\bd}
~, \quad 
  \\ & 
\{\cD_\a,\cD_\b\}=-4\bar{R}M_{\a\b}~,\quad
\{\cDB_\ad,\cDB_\bd\}=4R\bar{M}_{\ad\bd}~, 
\\
&{[}\cD_a,\cD_\b{]}=-\frac{\ri}{2}\bar{R}(\s_a)_{\b\gd}\cDB^{\gd}~,\qquad 
{[}\cD_a,\cDB_\bd{]}=\frac{\ri}{2}R(\s_a)_{\g\bd}\cD^{\g}~,   \\
&{[}\cD_a,\cD_b{]}=-|R|^2 M_{ab} ~, \label{1.2d}
\eea
\end{subequations}
with $R= {\rm const}$. The Riemann tensor of ${\rm AdS}^{4}$
may be deduced from \eqref{1.2d} to be  
\bea
{\mathfrak R}_{abcd} =-|R|^2  ( \eta_{ac}\eta_{bd} - \eta_{ad} \eta_{bc} )~.
\eea

The  three remaining superspaces are characterised by 
formally identical anti-commutation relations \cite{KT-M}
\begin{subequations}\label{RS^3}
\bea
&\{\cD_\a,\cD_\b\}= 0~, \quad \{\cDB_\ad,\cDB_\bd\}=0~,\quad
\{\cD_\a,\cDB_\bd\}=-2\ri\cD_{\a\bd}~, 
\\
&{[}\cD_\a,\cD_{\b\bd}{]}=\ri\ve_{\a\b}G^\g{}_{\bd}\cD_\g
~,\quad
{[}\cDB_\ad,\cD_{\b\bd}{]}=-\ri\ve_{\ad\bd}G_\b{}^\gd\cDB_\gd~,
 \\
&{[}\cD_{\a\ad},\cD_{\b\bd}{]}=
-\ri\ve_{\ad\bd}G_\b{}^\gd\cD_{\a\gd}
+\ri\ve_{\a\b}G^\g{}_\bd\cD_{\g\ad}~,
\eea
where  $G_{b} $ is covariantly constant,
\bea
\cD_A G_b = 0~.
\eea
\end{subequations}
The difference between these superspaces is encoded in the Lorentzian type of 
$G_a$.  Since $G^2 = G^a G_a $ is constant, the geometry 
\eqref{RS^3} describes  three different superspaces, 
${\mathbb M}^{4|4}_{T}$, ${\mathbb M}^{4|4}_{S}$ and ${\mathbb M}^{4|4}_{N}$, 
which correspond to the choices $G^2<0$, $G^2>0$ and $G^2=0$, 
respectively.
The Lorentzian manifolds, which are  the bosonic bodies of the superspaces 
${\mathbb M}^{4|4}_{T}$, ${\mathbb M}^{4|4}_{S}$ and ${\mathbb M}^{4|4}_{N}$,
are  ${\mathbb R}\times S^3$, 
${\rm AdS}_3 \times {\mathbb R}$ 
and a pp-wave spacetime,\footnote{The latter spacetime was shown in 
\cite{deMF-OS} to be isometric to the Nappi-Witten group NW$_4$ \cite{NappiW}.}   
respectively.
The Riemann curvature tensor of these spacetimes is 
\bea
{\mathfrak R}_{abcd} = &&\frac{1}{4} \Big\{ G_c (G_a \eta_{bd} - G_b \eta_{ad}) 
-  G_d (B_a \eta_{bc} - G_b \eta_{ac})  -G^2 (\eta_{ac}\eta_{bd} - \eta_{ad} \eta_{bc} )\Big\} ~.
\eea

The superspace ${\mathbb M}^{4|4}_{T}$ is the universal covering of 
$\cM^{4|4} =\sSU(2|1)$.  
The bosonic body of $\cM^{4|4}$ is $\sU(2)=(S^1 \times S^3)/{\mathbb Z}_2$.
The isometry group of  $\cM^{4|4}$ is $\sSU(2|1) \times \sU(2) $.
One can think of  ${\mathbb M}^{4|4}_{T}$ as a supersymmetric extension of 
Einstein's static universe. $\cN=1$ supersymmetric field theories 
on ${\mathbb R}\times S^3$ were studied in the mid-1980s  by Sen \cite{Sen}.
The superspace ${\mathbb M}^{4|4}_{S}$ is the universal covering of 
$\widetilde{\cM}^{4|4} =\sSU(1,1|1)$.
The bosonic body of $\widetilde{\cM}^{4|4}$ is 
$\sU(1,1)=({\rm AdS}_3 \times S^1)/{\mathbb Z}_2$.
 The isometry group of
$\widetilde{\cM}^{4|4}$ is $\sSU(1,1|1) \times \sU(2) $.

The superspace \eqref{AdS} is a maximally supersymmetric solution 
of anti-de Sitter supergravity described by the action (see, e.g., \cite{BK} for a review)
\bea
\label{omsg}
S_{\rm SUGRA} = - \frac{3}{ \k^2} 
\int {\rm d}^{4} x \rd^2\q\rd^2 \bar\q\,
E
+ \Big\{ \frac{ \m}{ \k^2} 
\int {\rm d}^{4} x \rd^2\q\,
\cal E  + {\rm c.c.} \Big\}
~, 
\eea
where $ \k$ is the gravitational coupling constant and  
$ \m$ a cosmological parameter. 
 The integration measures
 $E$ and $\cal E$ in \eqref{omsg}
 correspond to  the full superspace 
and its chiral subspace, respectively.  
The  equations of motion  corresponding to 
\eqref{omsg}
are 
\bea
G_a = 0~, \qquad R = \m~,
\eea
see \cite{BK} for a pedagogial derivation. 
Setting $\m =0$ in \eqref{omsg} gives the action 
for $\cN=1$ Poincar\'e supergravity \cite{WZ}.
Minkowski superspace \eqref{Minkowski} is a maximally supersymmetric solution of this theory. 

In this note, we are going to show that the superspaces \eqref{RS^3}
are maximally supersymmetric solutions 
of scale-invariant $R^2$ supergravity
\bea 
\label{1.8}
 S&=& \a \int \rd^4 x \,{\rm d}^2\q \rd^2 \bar \q\, E \, R \bar R 
+ \Big\{ \b \int \rd^4 x \,{\rm d}^2\q \, \cE \, R^3 +{\rm c.c.} \Big\}\non \\
&=&  \int \rd^4 x \,{\rm d}^2\q \rd^2 \bar \q\, E \, \Big\{\a R \bar R 
+ (\b R^2 + \bar \b \bar R^2) \Big\}~,
\eea
with $\a $ and $\b$ a real and a complex dimensionless parameters, respectively. 
The $\a$-term in \eqref{1.8} is generated as a one-loop quantum correction
in $\cN=1$ supersymmetric field theories coupled to supergravity 
\cite{BK86,BK88,BK88_2}. The component structure of this term was described in 
\cite{Theisen}. Although the $\b$-term in \eqref{1.8} breaks the $\sU(1)$ $R$-symmetry, 
adding such a contribution to the $\a$-term 
is completely natural, keeping in mind 
that a massless covariantly chiral scalar superfield $\F$, $\bar \cD_\ad \F=0$,  
is described in supergravity by an action 
\bea
S_{\rm matter} &=&  \int \rd^4 x \,{\rm d}^2\q \rd^2 \bar \q\, E \, \Big\{ \F \bar \F 
+ \hf  \x ( \F^2 +  \bar \F^2) \Big\}~,
\eea
with $\x$ a dimensionless parameter. The choice $\x=0$ corresponds to 
the conformal scalar multiplet model which is dual to the improved tensor multiplet 
\cite{deWR}.
Another natural  choice is $\x=1$ and corresponds to 
a non-conformal scalar multiplet which is dual to the
free tensor multiplet model \cite{Siegel}.

The higher-derivative supergravity model \eqref{1.8} has recently been studied  in 
\cite{Ferrara:2015ela}.\footnote{Action \eqref{1.8} can be rewritten in a manifestly super-Weyl invariant form, as in \cite{Ferrara:2015ela}, by introducing a chiral compensator 
$\f$, $\bar \cD_\ad \f=0$,  and replacing $R$ with 
the super-Weyl invariant chiral scalar 
${\mathbb R} = -\frac{1}{4} \f^{-2} (\bar \cD^2 - 4R) \bar \f$ 
and the full superspace measure $E$ with $E \, \f \bar \f$.  
Such a superconformal reformulation is sometimes useful, in particular for the component reduction, however it does not offer new insights to the analysis in this note.}
Along with the supergravity action, 
both terms in \eqref{1.8} have also been discussed in the framework of supersymmetric 
models for inflation, see  \cite{KS,FKVP} and references therein.

This note is organised as follows. In section 2 we briefly discuss the various superspace 
formulations for $\cN=1$ conformal supergravity, and the present a new 
derivation of the maximally supersymmetric backgrounds of off-shell $\cN=1$  supergravity. 
In section 3 we  prove that the curved superspaces described by \eqref{RS^3} are solutions 
of the no-scale supergravity model \eqref{1.8}. Some concluding comments are given in section 4. 


\section{A new derivation of the maximally supersymmetric backgrounds in off-shell
 $\cN=1$ supergravity} 

Every off-shell formulation for $\cN=1$ supergravity can be described using 
the superspace geometry pioneered by Howe thirty five years ago \cite{Howe} 
and soon after reviewed and further developed in \cite{GGRS}. 
This curved superspace geometry 
is based on the structure group $\sSL (2, {\mathbb C}) \times \sU(1) $,
and nowadays it is often referred to as $\sU(1)$ superspace.  
The algebra of supergravity covariant derivatives is as follows:
\begin{subequations}\label{Howe}
\bea
&& {} \qquad \{ \cD_\a , {\bar \cD}_\ad \} = -2{\rm i} \cD_{\a \ad} ~, 
 \\
\{\cD_\a, \cD_\b \} &=& -4{\bar R} M_{\a \b}~, \qquad
\{ {\bar \cD}_\ad, {\bar \cD}_\bd \} =  4R {\bar M}_{\ad \bd}~,  \label{Howe_b} \\
\left[ \cD_{\a} , \cD_{ \b \bd } \right]
     & = &
     {\rm i}
     {\ve}_{\a \b}
\Big({\bar R}\,\cDB_\bd + G^\g{}_\bd \cD_\g
- (\cD^\g G^\d{}_\bd)  M_{\g \d}
+2{\bar W}_\bd{}^{\gd \dot{\d}}
{\bar M}_{\gd \dot{\d} }  \Big) \non \\
&&
+ {\rm i} (\cDB_\bd {\bar R})  M_{\a \b}
-\frac{\ri}{3} \ve_{\a\b} \bar X^\gd \bar M_{\gd \bd} -\frac{\ri}{2} \ve_{\a\b} \bar X_\bd {\mathbb J}
~. 
\eea
\end{subequations}
Here the $\sU(1)_R$ generator ${\mathbb J}$ is normalised by
\begin{align}
[{\mathbb J} , \cD_\alpha] = \cD_\alpha~,\qquad 
[{\mathbb J} , \bar \cD_\dalpha] = -\bar \cD_\dalpha~.
\label{2.22}
\end{align}
The torsion superfields $R$, $G_{\alpha \dalpha}$, $W_{\alpha \beta \gamma}$,
and $X_\alpha$ obey the Bianchi identities:
\begin{subequations}
\begin{gather}
\bar \cD_\ad R = 0~, \qquad
\bar \cD_\dalpha X_\alpha = 0~, \qquad
\bar \cD_\dalpha W_{\alpha \beta \gamma} = 0  ~, \\
X_\alpha = \CD_\alpha R - \bar \cD^\dalpha G_{\alpha \dalpha}~, \qquad
\cD^\alpha X_\alpha = \bar \cD_\dalpha \bar X^\dalpha~. 
\end{gather}
\end{subequations}
The reason why the superspace geometry defined by \eqref{Howe} is adequate 
to describe $\cN=1$ conformal supergravity is the fact that the 
the algebra \eqref{Howe} does not change under a 
super-Weyl transformation 
\bea
\cD_\alpha' &= \re^{\hf L}
	\Big(\cD_\alpha + 2 (\cD^\beta L) \,M_{\beta \alpha}
	- \frac{3}{2} (\cD_\alpha ) L \, {\mathbb J}\Big)~
\label{super-Weyl-Howe}
\eea
accompanied by induced transformations of the torsion superfields.
The parameter $L$ in \eqref{super-Weyl-Howe} is a real unconstrained superfield. 

Before turning to the derivation of the maximally supersymmetric backgrounds
of  supergravity, it is worth commenting on other superspace approaches to describe 
$\cN=1$ conformal supergravity. 
The $\sU(1)$ superspace of \cite{Howe} is a gauge fixed version of 4D $\cN=1$ 
conformal superspace \cite{Butter4DN=1}, in which the entire superconformal algebra
$\sSU(2,2|1)$ 
is gauged in superspace (see also  \cite{ButterK} for a review of the relationship 
between the $\sU(1) $ and conformal superspaces).  When studying 
supersymmetric backgrounds of supergravity, it suffices to work with 
$\sU(1)$ superspace, and therefore we do not use conformal superspace 
in this note. 

The superspace geometry developed by Grimm, Wess and Zumino \cite{GWZ}
is obtained from \eqref{Howe} by setting 
\bea
X_\a =0~.
\label{2.5}
\eea
Under this condition, the  $\sU(1)_R$ connection can be gauged away 
and the structure group reduces to $\sSL (2, {\mathbb C})$.
 Requirement \eqref{2.5} can always be achieved by applying a specially chosen 
 super-Weyl transformation \eqref{super-Weyl-Howe}.
If such a super-Weyl gauge is chosen, one stays with a 
residual super-Weyl plus $\sU(1)$ gauge freedom given by \cite{HT}
\bea
\cD'_\a = {\rm e}^{ \bar \s - \s/2  } \Big(
\cD_\a + (\cD^\b \s) \, M_{\a \b} \Big)~, \qquad \bar \cD_\ad \s =0~.
\eea

As is well known (see, e.g.,  \cite{GGRS} for a review),
the different off-shell formulations for $\cN=1$ supergravity are obtained 
by coupling conformal supergravity (described, e.g. using  $\sU(1)$ superspace)
to a compensator. 
The latter is a chiral scalar in the case of the old minimal formulation \cite{WZ,old}, 
a real linear superfield for the new minimal formulation \cite{new}, 
and a complex linear superfield for the non-minimal formulation \cite{non-min,SG}. 
Our analysis of maximally supersymmetric backgrounds of supergravity 
does not require fixing any specific compensator.

We now recall an important  theorem concerning the maximally supersymmetric 
backgrounds \cite{KNT-M,K15Corfu}. 
For any (off-shell) supergravity theory in $D$ dimensions, 
all maximally supersymmetric spacetimes correspond 
to those  supergravity backgrounds which 
are characterised by the following properties:
(i) 
all Grassmann-odd components 
of the superspace torsion and curvature tensors vanish; and 
(ii) 
all Grassmann-even components of the torsion and curvature tensors are annihilated 
by the spinor derivatives. 
In the case of 4D $\cN=1$ supergravity, this theorem means the following:
\begin{subequations}
 \bea
 X_\a &=& 0~;  \label{2.7a}\\
 W_{\a\b\g}&=&0 ~; \label{2.7b} \\
 \cD_\a R =0 ~\longrightarrow ~\cD_A R &=&0 ~; \\
 \cD_\a G_{\b\bd} = \bar \cD_\ad G_{\b\bd} =0~
\longrightarrow ~\cD_A G_{\b\bd} &=&0~. \label{2.5c}
\eea
\end{subequations}
Equation \eqref{2.5c} has an integrability condition that follows from \eqref{Howe_b}.
It is 
\bea
0 = \{ \bar \cD_\ad , \bar \cD_\bd \} G_{\g \gd} = 4R \bar M_{\ad \bd} G_{\g \gd} 
= 2R (\ve_{\gd \ad} G_{\g \bd} + \ve_{\gd \bd} G_{\g \ad})~,
\eea
and therefore 
\bea
R G_{\a\ad} =0~. 
\label{2.7}
\eea
Eq. \eqref{2.7a} tells us that all maximally supersymmetric backgrounds
 are realised in terms of  the 
 Grimm-Wess-Zumino superspace geometry \cite{GWZ}.

Relation \eqref{2.7} (actually its $\q$-indepdentent part)
was given in \cite{FS} without derivation.
Let us also show that \eqref{2.7} is a simple consequence of 
the general analysis given in section 6.4 of  \cite{BK}.
Consider a background superspace  $(\cM^{4 |4}, \cD)$. 
A supervector field $\x= \x^B E_B=\x^bE_b + \x^\b E_\b + \bar \x_\bd \bar E^\bd$  on $(\cM^{4 |4}, \cD)$ is called  Killing if 
\bea
\d_\cK \cD_A = [\cK , \cD_A] =0 ~, \qquad 
\qquad \cK:= \x^B (z) \cD_B + \hf K^{bc} (z) M_{bc} + \ri \t (z) {\mathbb J}~, 
\eea
for some Lorentz ($K^{bc}$) and $R$-symmetry ($\t$) parameters. 
All parameters  $\x^\b$, $K^{bc}$,  $\t$ are determined in terms of $\x^b$,

Let $\x= \x^A E_A $ be a conformal Killing supervector field of  $(\cM^{4 |4}, \cD)$. 
As demonstrated in section 6.4 of  
 \cite{BK},  its explicit form is 
 \bea
\x^A = \Big( \x^a, \x^\a, \bar \x_\ad \Big) 
= \Big( \x^a, -\frac{\ri }{8} \bar \cD_\bd \x^{\bd \a} , 
-\frac{\ri}{8} \cD^\b \x_{\b \ad} \Big)~,
\eea
where the vector component $\x_{\a\ad} $ is real and obeys the equation \cite{BK}
\bea
\cD_{(\a} \x_{\b) \bd} =0  ~,
\label{2.10}
\eea
which implies 
\bea
(\cD^2 +2 \bar R ) \x_{\a\ad}=0~.
\label{2.11}
\eea
In accordance with \eqref{2.5c}, $G_{\a\ad}$ is covariantly constant, and 
hence it is a solution of \eqref{2.10}. Then \eqref{2.11} reduces to \eqref{2.7}. 


\section{Maximally supersymmetric solutions of no-scale $R^2$ supergravity}

We now prove that the curved superspaces described by \eqref{RS^3} are solutions 
of the no-scale supergravity model \eqref{1.8}. For this we will use 
the background-field method for $\cN=1$ supergravity 
as developed by Grisaru and Siegel \cite{GrisaruSiegel} and further elaborated in \cite{BK}.
 
We denote  infinitesimal increments of the supergravity
prepotentials by $H^{a} $ and $\s$, 
where $H^a$ is real unconstrained and  $\s$ is covariantly chiral, 
$\bar \cD_\ad \s=0$. The variations of various supergravity functionals under such 
an infinitesimal change in the prepotentials was computed  
in section 5.6 of the book \cite{BK} (see also \cite{BK88}).
The results we need here are: 
\begin{subequations}
\bea
&& \d \int \rd^4 x \,{\rm d}^2\q \rd^2 \bar \q\, E \, R \bar R 
= - \frac{1}{4} \int \rd^4 x \,{\rm d}^2\q \rd^2 \bar \q\, E \, \Big\{ \s \cD^2 R + \bar \s \bar \cD^2 \bar R\Big\} \non \\
&& +\hf \int \rd^4 x \,{\rm d}^2\q \rd^2 \bar \q\, E \, H^{\a\ad}
\Big\{2 R \bar R G_{\a\ad} 
- \frac{1}{6} (\cD^2 R + \bar \cD^2 \bar R) 
+\frac{\ri} {6} \cD_{\a\ad} ( \bar \cD^2 \bar R - \cD^2 R ) \non \\
&&\qquad \qquad \qquad \qquad \qquad \quad
+ \frac{2}{3}R \stackrel{\leftrightarrow}{D}_{\a\ad}  \bar R
+\frac{1}{3} (\cD_\a R) \bar \cD_\ad \bar R \Big\} ~, \label{3.1a}
\\
&& \d \int \rd^4 x \,{\rm d}^2\q \rd^2 \bar \q\, E \, R^2
= 3\int \rd^4 x \,{\rm d}^2\q \rd^2 \bar \q\, E \, ( \s -\bar \s) R^2 \non \\
&& + \int \rd^4 x \,{\rm d}^2\q \rd^2 \bar \q\, E \, H^{\a\ad} \Big\{
G_{\a\ad} - \ri \cD_{\a\ad} \Big\} R^2~. \label{3.1b}
\eea
\end{subequations}
It is seen that both variations \eqref{3.1b} and \eqref{3.1b}
vanish for the backgrounds  \eqref{RS^3}.
If the parameter $\b$ in  \eqref{1.8} is non-zero, $\b \neq 0$, 
the anti-de Sitter superspace \eqref{AdS} is not a solution of the equations of motion 
for \eqref{1.8}.

In accordance with \eqref{2.7b}, 
all maximally supersymmetric backgrounds of $\cN=1$ supergravity
are conformally flat.\footnote{This is not true for some  maximally supersymmetric backgrounds 
of $\cN=2$ supergravity \cite{BIL}.}
Therefore all of them are solutions of the equations of motion for 
$\cN=1$ conformal supergravity described by the chiral action \cite{Siegel78,Zumino}
\bea 
I_{\rm CSG} =  \int \rd^4x\, \rd^2\q\, \cE\,  W^{\a\b \g}W_{\a\b\g} 
~+~{\rm c.c.} 
\non
\eea


\section{Concluding comments}

It is instructive to compare the maximally supersymmetric backgrounds
\eqref{AdS} and \eqref{RS^3} with their counterparts for three-dimensional $\cN=2$ supergravity. 

In three dimensions, the maximally supersymmetric backgrounds of 
off-shell $\cN=2$ supergravity were classified in \cite{KLRST-M}, 
and also reviewed and elaborated in \cite{K15Corfu}.
The  three-dimensional analogue of  \eqref{AdS}
is the (1,1) AdS superspace \cite{KT-M11}. 
The three-dimensional analogues of the backgrounds \eqref{RS^3}
are given by the following algebra of covariant derivatives 
$\cD_A = (\cD_a, \cD_\a , \bar \cD^\a)$
\begin{subequations}\label{4.39}
\bea
\{\cD_\a,\cD_\b\}
&=&
0
~,\qquad
\{\cDB_\a,\cDB_\b\}
=
0~,
\\
\{\cD_\a,\cDB_\b\}
&=&
-2 \ri (\g^c)_{\a\b} \Big(  \cD_c - 2\cS M_c
-\ri  \cC_{c} {\mathbb J} \Big)
+4\ve_{\a\b}\Big( \cC^{c}M_{c}- \ri \cS {\mathbb J}\Big)
~, \\
{[}\cD_{a},\cD_\b{]}
&=&
\ri\ve_{abc}(\g^b)_\b{}^{\g}\cC^c\cD_{\g}
+ (\g_a)_\b{}^\g \cS \cD_{\g}~, \\
{[}\cD_{a},\cDB_\b{]}
&=&
-\ri\ve_{abc}(\g^b)_\b{}^{\g}\cC^c\cDB_{\g}
+(\g_a)_\b{}^{\g}\cS \bar \cD_{\g}~, \\
{[}\cD_a,\cD_b]{}
&=&4  \ve_{abc}\Big( \cC^c \cC_d
+\d^c{}_d \cS^2
\Big)M^d ~.
\eea
\end{subequations}
Here $M_c$ denotes the Lorentz generator (defined in  \cite{KLRST-M}) 
and the  $\sU(1)_R$ generator ${\mathbb J}$ is defined similarly to \eqref{2.22}.
The scalar $\cS$ and vector $\cG_b$ 
components of the torsion tensor are constrained by 
\bea
{\cD}_A  {\cS}= 0~,
\qquad \cD_\a \cC_b =0 \quad \Longrightarrow \quad
{\cD}_{a} {\cC}_b=
2\ve_{abc}{\cC}^c {\cS} ~,
\eea
and hence $ \cC^b \cC_b ={\rm const}$. 
We point out that the solution with ${\cC}_a =0$ corresponds to the  (2,0) AdS superspace
\cite{KT-M11}. However, here we are interested in the case $\cC_b \neq 0$. 
When both $\cS$ and $\cC_b$ are non-vanishing, the above curved superspace is 
a maximally supersymmetric solution of topologically massive type II supergravity 
 \cite{K15Corfu}. In the case $\cS=0$ and $\cC_b \neq 0$, the above superspace is  
a solution of three-dimensional $R^2$ supergravity \cite{KNT-M15}.
 
 One of the most interesting properties of the maximally supersymmetric backgrounds
 \eqref{RS^3} is that they allow for  the Maxwell-Goldstone multiplet models 
which describe partial $\cN=2 \to \cN=1$ supersymmetry breaking \cite{KT-M} 
and reduce to the Bagger-Galperin model  \cite{BG} in the flat limit, $G_a \to 0$.
 
The $\cN=2$ analogue of  scale-invariant $R^2$ supergravity
\eqref{1.8} was given in \cite{KN15}. It is of interest to see which rigid $\cN=2$ maximally supersymmetric backgrounds \cite{BIL} are solutions of this theory.  
\\

\noindent
{\bf Acknowledgements:}\\
It is my pleasure to acknowledge the hospitality of Dima Sorokin 
and the INFN, Sezione di Padova, where this project was designed.  
I also thank Luca Martucci for asking a question that 
provided the rationale for writing up
the construction described in this paper. 
Daniel Butter is gratefully acknowledged for helpful correspondence.  
Joseph Novak is gratefully acknowledged for comments on the manuscript. 
This work is supported in part by the Australian Research Council,
project No. DP160103633. 

\begin{footnotesize}

\end{footnotesize}

\end{document}